\documentclass[aps,prd,twocolumn,showpacs,preprintnumbers,nofootinbib,amsmath,amssymb]{revtex4}

\usepackage{graphicx}
\usepackage{dcolumn}
\usepackage{bm}
\usepackage{amsmath}
\usepackage[normalem]{ulem}

\newcommand{\beq}{\begin{eqnarray}}
\newcommand{\eeq}{\end{eqnarray}}

\newcommand{\Tr}{\ensuremath{\mathrm{Tr}}}

\newcommand{\Dmrns}{{\cal D}_{\mu\rho,\nu\sigma}}

\def\spose#1{\hbox to 0pt{#1\hss}}
\def\ltapprox{\mathrel{\spose{\lower 3pt\hbox{$\mathchar"218$}}
 \raise 2.0pt\hbox{$\mathchar"13C$}}}

\begin{document}

\title{Gauge-invariant field-strength correlators for QCD in a magnetic background}
\author{Massimo D'Elia}
\email{massimo.delia@unipi.it}
\affiliation{
Dipartimento di Fisica dell'Universit\`a
di Pisa and INFN - Sezione di Pisa,\\ Largo Pontecorvo 3, I-56127 Pisa, Italy}

\author{Enrico Meggiolaro}
\email{enrico.meggiolaro@unipi.it}
\affiliation{
Dipartimento di Fisica dell'Universit\`a
di Pisa and INFN - Sezione di Pisa,\\ Largo Pontecorvo 3, I-56127 Pisa, Italy}

\author{Michele Mesiti}
\email{mesiti@pi.infn.it}
\affiliation{
Dipartimento di Fisica dell'Universit\`a
di Pisa and INFN - Sezione di Pisa,\\ Largo Pontecorvo 3, I-56127 Pisa, Italy}

\author{Francesco Negro}
\email{fnegro@pi.infn.it}
\affiliation{
Dipartimento di Fisica dell'Universit\`a
di Pisa and INFN - Sezione di Pisa,\\ Largo Pontecorvo 3, I-56127 Pisa, Italy}

\date{\today}

\begin{abstract}
We consider the properties of 
the gauge-invariant two-point correlation functions of the gauge-field
strengths for QCD in the presence of a magnetic background field at zero temperature. 
We discuss the general structure of
the correlators in this case
and provide the results of an exploratory lattice study for $N_f = 2$ QCD
discretized with unimproved staggered fermions. Our analysis provides evidence
for the emergence of anisotropies in the non-perturbative part of the
correlators and for an increase of the gluon condensate
as a function of
the external magnetic field.
\end{abstract}

\maketitle

\section{Introduction}
\label{intro}

The study of strong interactions in the presence of strong magnetic
fields has attracted an increasing interest in the last few 
years~(see, e.g.,~Ref.~\cite{lecnotmag}).
This is justified by the great phenomenological relevance of the 
issue:
the physics of 
some compact astrophysical
objects, like magnetars~\cite{magnetars}, 
of non-central heavy ion collisions~\cite{hi1,hi2,hi3,hi4, tuchin} 
and of the early Universe~\cite{vacha,grarub}
involve the properties
of quarks and gluons in the presence of magnetic backgrounds
going from $10^{10}$ Tesla up to $10^{16}$ 
Tesla ($|e| B \sim 1$ GeV$^2$).
The study is also interesting from a purely theoretical point
of view, since it reveals new non-perturbative 
features of non-Abelian gauge theories.

One emerging feature
is that gluon fields, even if not directly coupled to electromagnetic 
fields, can be significantly affected 
by them. This is not unexpected, since  
effective QED-QCD interactions, induced by quark loop contributions,
can be important, because of the non-perturbative nature of the 
theory~\cite{anisotropic, chernodub, musak, elze1, elze2, mueller, galilo, KojoSu1, 
KojoSu2, watson, andersen, ozaki, kamikado, mueller2, simonov2, 
Simonov:2015yka,Cao:2015cka}.
In particular, a uniform magnetic background is expected
to lead to gluon-field anisotropies~\cite{anisotropic, chernodub, galilo, ozaki},
which may also have phenomenological implications.
This has been confirmed
by various lattice QCD 
studies~\cite{demusa,DEN,reg0,Ilgenfritz:2012fw,reg2,kovacs,EB,Ilgenfritz:2013ara}. In particular, anisotropies have been observed in various
pure gauge quantities, like the average plaquettes 
taken in different planes~\cite{Ilgenfritz:2012fw, reg2}
and the static quark-antiquark potential ~\cite{struzzo},
with possible effects on the spectra of heavy quark bound states~\cite{calimero}.

In the present study, looking for a more systematic analysis
of the non-perturbative vacuum modifications induced by the magnetic
field, we consider the gauge-invariant two-point field-strength correlators
(see Ref. \cite{DDSS02} for a complete review on this subject),
which are defined as
\beq
\Dmrns(x) = g^2 \langle
\Tr [ G_{\mu\rho}(0) S(0,x) G_{\nu\sigma}(x) S^\dagger(0,x) ]
\rangle ,
\label{defcorr}
\eeq
where $G_{\mu\rho} = T^aG^a_{\mu\rho}$ is the field-strength tensor,
$T^a$ are the $SU(3)$ generators 
in the fundamental representation, and $S(0,x)$
is the parallel transport from $0$ to $x$ along a straight line
(Schwinger line),
which is needed to make the correlator gauge invariant.
Such correlators have been first considered,
together with analogous correlators involving fermionic
fields, to take into account the non-uniform
distributions of the vacuum 
condensates~\cite{Gromes:1982su, Campostrini:1984xr, CDO1986, STY1995}. 
Then, they have been widely used to parametrize the non-perturbative
properties of the QCD vacuum, especially within the framework of the so-called
{\it Stochastic Vacuum Model}~\cite{Dosch87,Dosch-Simonov88,Simonov88},
since they represent the leading (Gaussian) term in the cumulant expansion.
Gluon-field correlators have been also exploited
as a tool to explore the response of the QCD vacuum to
external magnetic fields~\cite{Simonov:2015yka,Simonov:2015xta}, however
this has usually been done by considering the unmodified 
correlators, i.e. computed at zero magnetic field. 
The question that we approach here is 
different: how are the zero-temperature correlators themselves modified by the
background field?

A first issue to be considered regards the symmetry properties
and the associated parametrization of the correlators.
In the vacuum and in the absence of external sources, Lorentz symmetry
implies a simple form for the two-point functions in Eq.~(\ref{defcorr}), 
which can be expressed in terms of two independent scalar functions
of $x^2$, which are usually denoted by ${\cal D} (x^2)$ and
${\cal D}_1 (x^2)$~\cite{Dosch87,Dosch-Simonov88,Simonov88}:
\beq
\label{usual_param}
\lefteqn{
\Dmrns(x) =
(\delta_{\mu\nu}\delta_{\rho\sigma} - \delta_{\mu\sigma}\delta_{\rho\nu})
\left[ {\cal D}(x^2) + {\cal D}_1(x^2) \right] } \\
& & + (x_\mu x_\nu \delta_{\rho\sigma} - x_\mu x_\sigma \delta_{\rho\nu}
+ x_\rho x_\sigma \delta_{\mu\nu} - x_\rho x_\nu \delta_{\mu\sigma})
\frac{\partial{\cal D}_1(x^2)}{\partial x^2} . \nonumber
\eeq
The presence of a uniform external field breaks Lorentz symmetry
explicitly, so that such a parametrization 
is not justified any more, especially 
in the light of the already observed propagation of symmetry breaking
from the electromagnetic to the gluon sector.
Therefore in Section~\ref{structure}, on the 
basis of 
the residual symmetries of the theory, we discuss 
what is the form the correlators can take 
in this case.

Then, in Section~\ref{lattice}, we present an exploratory
lattice determination of the gluon-field correlators
in a magnetic field which
is performed for $N_f = 2$ QCD, discretized by means of unimproved 
staggered fermions, and exploits {\it cooling} as a 
technique to smooth out ultraviolet fluctuations.
Numerical results will be analyzed, within the parametrization
proposed in Section~\ref{structure}, in order
to clarify which properties of the gluon correlators
are mostly affected by the presence of the magnetic field.
The analysis will then focus on a quantity of phenomenological 
interest which can be extracted from the gluon correlators,
the so-called {\it gluon condensate}.
Finally, in Section~\ref{finally}, we draw our conclusions.

\section{Field correlators in a constant magnetic background}
\label{structure}

As we have already emphasized 
in the Introduction, the presence of an external field
breaks Lorentz symmetry ($SO(4)$ symmetry in the Euclidean space), so that 
the most general parametrization is more complex than the one 
reported in Eq.~\eqref{usual_param}. Let us discuss this point in more detail.

The required general symmetry properties for the correlator in 
Eq.~(\ref{defcorr}) are that it be left unchanged 
under exchange of the $\mu \rho$ and $\nu \sigma$ pairs,
and antisymmetric both in the $\mu \rho$ and in the $\nu \sigma$
indices, i.e.~we can write in general
\beq
\Dmrns = \sum_{n} f_n T^{(n)}_{\mu\rho,\nu\sigma},
\label{general_param}
\eeq
where:\\
i) $T^{(n)}_{\nu\sigma,\mu\rho} = T^{(n)}_{\mu\rho,\nu\sigma}$, and
ii) $T^{(n)}_{\rho\mu,\nu\sigma} = T^{(n)}_{\mu\rho,\sigma\nu} =
-T^{(n)}_{\mu\rho,\nu\sigma}$.
\\
 A class of tensors satisfying such properties 
is written as
\beq
T^{(A,B)}_{\mu\rho,\nu\sigma} \equiv A_{\mu\nu}B_{\rho\sigma} -
A_{\rho\nu}B_{\mu\sigma} - A_{\mu\sigma}B_{\rho\nu} +
A_{\rho\sigma}B_{\mu\nu},
\label{T^AB}
\eeq
where $A_{\mu\nu}$ and $B_{\mu\nu}$ are two rank-2 tensors which are both
symmetric ($A_{\nu\mu}=A_{\mu\nu}$, $B_{\nu\mu}=B_{\mu\nu}$) or both
antisymmetric ($A_{\nu\mu}=-A_{\mu\nu}$, $B_{\nu\mu}=-B_{\mu\nu}$).

In the
absence of external background fields, there are only two available 
rank-2 tensors, which are symmetric and 
can be constructed in terms of the metric tensor
$g_{\mu\nu} = \delta_{\mu\nu}$ and of the four-vector $x_\mu$:
they are $\delta_{\mu\nu}$ itself and $x_\mu x_\nu$,
so that the most general parametrization reads 
as in Eq.~(\ref{usual_param}):
\beq
\Dmrns = f_1 T^{(1)}_{\mu\rho,\nu\sigma}
+ f_2 T^{(2)}_{\mu\rho,\nu\sigma},
\label{usual_param_bis}
\eeq
where
\beq
T^{(1)}_{\mu\rho,\nu\sigma} \equiv&
\frac{1}{2} T^{(\delta,\delta)}_{\mu\rho,\nu\sigma} &=~
\delta_{\mu\nu}\delta_{\rho\sigma} - \delta_{\mu\sigma}\delta_{\rho\nu},
\nonumber\\
T^{(2)}_{\mu\rho,\nu\sigma} \equiv& T^{(xx,\delta)}_{\mu\rho,\nu\sigma} &=~
x_\mu x_\nu \delta_{\rho\sigma} - x_\mu x_\sigma \delta_{\rho\nu}
\nonumber\\
& &+~~ x_\rho x_\sigma \delta_{\mu\nu} - x_\rho x_\nu \delta_{\mu\sigma},
\label{T1&T2}
\eeq
($T^{(xx,xx)} = 0$ by construction), while $f_1 \equiv {\cal D} + {\cal D}_1$ and $f_2 \equiv 
\frac{\partial{\cal D}_1}{\partial x^2}$ (or, equivalently, ${\cal D}$ and
${\cal D}_1$) are two scalar functions of $x^2$.

In the presence of an external background 
field $F_{\mu\nu}$, instead, many additional 
rank-2 tensors appear, both antisymmetric
(like $F_{\mu\nu}$ itself or $H_{\mu\nu} \equiv h_\mu x_\nu - h_\nu x_\mu$,
where $h_\mu \equiv F_{\mu\nu}x_\nu$) and
symmetric (like $F^{(2)}_{\mu\nu} \equiv F_{\mu\alpha}
F_{\alpha\nu}$ or 
$M_{\mu\nu} \equiv p_\mu x_\nu + p_\nu x_\mu$,
where $p_\mu \equiv F^{(2)}_{\mu\nu}x_\nu = F_{\mu\alpha}h_\alpha$).
Correspondingly, many more terms appear in the
parametrization in Eq.~(\ref{general_param}),
with new rank-4 tensors like
$\frac{1}{2} T^{(F,F)}_{\mu\rho,\nu\sigma}$,
$T^{(F,H)}$, 
$T^{(\delta,F^{(2)})}$, 
$T^{(xx,F^{(2)})}$, $T^{(\delta,hh)}$, 
$T^{(\delta,M)}$ and so on.
Moreover, for a magnetic field directed along the 
$z$ axis, the coefficients $f_n$ will depend separately 
on $x^2 + y^2$ and on $z^2 + t^2$, because of the breaking of
the Euclidean $SO(4)$ symmetry.
All that makes a numerical analysis based on the more general parametrization
\eqref{general_param} of the correlator quite involved
and not easily affordable.

On the other hand, in our present investigation on the lattice,
we shall consider only correlators of the kind
\beq
\mathcal{D}_{\mu\nu,\xi}(d) \equiv \mathcal{D}_{\mu\nu,\mu\nu}(x = d\hat{\xi}),
\label{latcorr_measured}
\eeq
where the two plaquettes are parallel to each other and the separation $x$ is
along one ($\hat{\xi}$) of the four basis vectors of the lattice
[$\hat{x}=(1,0,0,0)$, $\hat{y}=(0,1,0,0)$, $\hat{z}=(0,0,1,0)$,
$\hat{t}=(0,0,0,1)$].
These amount, in general, to $24$ different correlation functions. Without any 
additional external field, the symmetries of the system group these $24$
correlators into two equivalence classes,
usually denoted (as in \cite{DP1992})
$\mathcal{D}_\parallel$ (when $\xi=\mu$ or $\xi=\nu$ ) and
$\mathcal{D}_\perp$ (when $\xi\neq\mu$ and $\xi\neq\nu$), with
\beq
\label{Dpara-Dperp}
&&{\cal D}_\parallel = {\cal D} + {\cal D}_1
+ x^2 \frac{\partial {\cal D}_1}{\partial x^2} ,\nonumber\\
&&{\cal D}_\perp = {\cal D} + {\cal D}_1 .
\eeq
The dependence of the correlators on the distance $d$, in the case of zero
external field, has been { directly determined by numerical simulations
on the lattice in} \cite{DP1992,npb97,plb97,EM99,DDM2003}:\footnote{In another
approach \cite{BBV98}, the correlators have been extracted from lattice
calculations of the heavy-quark potential, by analysing field-strength
insertions into a Wilson loop under the assumption of factorisation, as in the
{\it Stochastic Vacuum Model}.}
the functions $\mathcal{D}$ and $\mathcal{D}_1$ have been parametrized
in the form
\beq
\label{param-D-D1}
\mathcal{D} = \frac{a_0}{d^4} + A_0 e^{-\mu d} ,\nonumber\\ 
\mathcal{D}_1 = \frac{a_1}{d^4} + A_1 e^{-\mu d} ,
\eeq
where the terms $\sim 1/d^4$ are of perturbative origin and are necessary
to describe the short distance behavior of the correlators,
while the exponential terms represent the non-perturbative contributions.
In particular, the coefficients $A_0$ and $A_1$ can be directly linked to the
{\it gluon condensate} of the QCD vacuum (see Eq. \eqref{g2def} below),
while the correlation length $1/\mu$, which sets the scale
of the spatial variations of the non-perturbative vacuum fluctuations,
governs the effect of the condensate on the levels of $Q\bar{Q}$ bound
states~\cite{Gromes:1982su, Campostrini:1984xr, CDO1986, STY1995},
and, moreover, it enters the description of various
QCD vacuum models~\cite{Dosch87,Dosch-Simonov88,Simonov88}.
In Refs.~\cite{npb97,plb97}, the perturbative-like terms had the form
$\sim e^{-\mu_a d}/d^4$: in the present work, instead (following what was done
also in Ref.~\cite{DDM2003}), we have neglected the exponential term $e^{-\mu_a d}$,
by fixing $\mu_a = 0$, since, in the spirit of the {\it Operator Product
Expansion}, we will concentrate on the behavior of the correlators at
short distances.
In terms of the quantities ${\cal D}_\parallel$ and ${\cal D}_\perp$
defined in Eq.~\eqref{Dpara-Dperp}, the parametrization
\eqref{param-D-D1} reads:
\beq
\label{param-Dpara-Dperp}
&&\mathcal{D}_\parallel = \left[ A_0+ A_1 \left(1-\frac{1}{2}\mu d\right)
\right] e^{-\mu d} + \frac{a_0-a_1 }{d^4} ,\nonumber\\
&&\mathcal{D}_\perp =  \left(A_0+ A_1\right) e^{-\mu d} + \frac{a_0+a_1}{d^4} \, .
\eeq

In the presence of a constant and uniform 
magnetic field $\vec{B}$ oriented along the
$z$ axis, i.e. $F_{xy} \neq 0$, the $SO(4)$ Euclidean symmetry
breaks into
\beq
SO(2)_{xy} \otimes SO(2)_{zt}.
\label{res_cont_symmetry}
\eeq
By virtue of this residual symmetry (which implies two equivalence relations,
one between the two {\it transverse} directions $\hat{x} \sim \hat{y}$
and another between the two {\it longitudinal} [or: ``{\it parallel}'']
directions $\hat{z} \sim \hat{t}$),
the 24 correlation functions in Eq.~(\ref{latcorr_measured}) are grouped
into $8$ equivalence classes, as shown in Table~\ref{table:corrclasses}.
\begin{table}[h]
\begin{tabular}{| c | c |}
\hline \textbf{Class Name} & \textbf{Elements} $(\mu\nu,\xi)$ \\
\hline $\mathcal{D}^{tt,t}_\parallel$ & (12,1) , (12,2)\\
\hline $\mathcal{D}^{tt,p}_\perp$     & (12,3) , (12,4)\\
\hline $\mathcal{D}^{tp,t}_\parallel$ & (13,1) , (14,1) , (23,2) , (24,2)\\
\hline $\mathcal{D}^{tp,p}_\parallel$ & (13,3) , (14,4) , (23,3) , (24,4)\\
\hline $\mathcal{D}^{tp,t}_\perp$     & (13,2) , (14,2) , (23,1) , (24,1)\\
\hline $\mathcal{D}^{tp,p}_\perp$     & (13,4) , (14,3) , (23,4) , (24,3)\\
\hline $\mathcal{D}^{pp,t}_\perp$     & (34,1) , (34,2) \\
\hline $\mathcal{D}^{pp,p}_\parallel$  & (34,3) , (34,4) \\
\hline
\end{tabular}
\caption{The $8$ equivalence classes of linearly independent correlation
functions in which the 24 components of the correlator
$\mathcal{D}_{\mu\nu,\xi}$, defined in 
Eq.~(\ref{latcorr_measured}), can be grouped.
The superscripts $t$, $p$ stand respectively for the $\hat{x},\hat{y}$
(\emph{transverse} to $\vec{B}$) directions and for the $\hat{z},\hat{t}$
(``\emph{parallel}'' to $\vec{B}$) directions.}
\label{table:corrclasses}
\end{table}
It must be noted that we can also group the $8$ correlation functions in Table 
\ref{table:corrclasses} into $3$ ``over-classes'', by the plaquette indices
$\mu$ and $\nu$:
\beq
&&\mathcal{D}^{tt} = \{\mathcal{D}^{tt,t}_\parallel,\mathcal{D}^{tt,p}_\perp\},
\nonumber \\
&&\mathcal{D}^{tp} = \{\mathcal{D}^{tp,t}_\parallel,
\mathcal{D}^{tp,p}_\parallel,\mathcal{D}^{tp,t}_\perp,
\mathcal{D}^{tp,p}_\perp\}, \label{overclasses}\\
&&\mathcal{D}^{pp} = \{\mathcal{D}^{pp,t}_\perp,\mathcal{D}^{pp,p}_\parallel\}.
\nonumber 
\eeq
In other words, the 24 correlation functions in Eq. (\ref{latcorr_measured}) can be written,
using the parametrization in Eq. (\ref{general_param}), as linear combinations of eight
linearly independent tensors $T^{(n)}$, e.g., the two fundamental tensors
in Eq. (\ref{T1&T2}) {\it plus} six other linearly independent tensors among those
listed below Eq. (\ref{T1&T2}), with eight (non-zero) independent functions $f_n$.
(For these particular correlators, the contribution from any other possible
tensor will be i) simply zero, or ii) a linear combination of the first eight
independent tensors.)
Now the question is: how should we parametrize the eight independent functions
$f_n$, or, equivalently, their eight linear comnbinations which represent
the eight functions listed in the first column of Table \ref{table:corrclasses}?
We shall use for these eight functions the following parametrization
(which is a simple generalization of the parametrization (\ref{param-Dpara-Dperp}) used in the
$B=0$ case):
\beq
\label{newtradpar}
&&\mathcal{D}^{tt,t}_\parallel =
\left[ A^{tt}_0+ A^{tt}_1 \left(1-\frac{1}{2}\mu^{tt,t}d\right) \right]
e^{-\mu^{tt,t}d} + \frac{a^{tt,t}_\parallel}{d^4} ,\nonumber\\
&&\mathcal{D}^{tt,p}_\perp =
(A^{tt}_0+ A^{tt}_1) e^{-\mu^{tt,p}d} + \frac{a^{tt,p}_\perp}{d^4} ,\nonumber\\
&&\mathcal{D}^{tp,t}_\parallel =
\left[ A^{tp}_0+ A^{tp}_1 \left(1-\frac{1}{2}\mu^{tp,t}d\right) \right]
e^{-\mu^{tp,t}d} + \frac{a^{tp,t}_\parallel}{d^4} ,\nonumber\\
&&\mathcal{D}^{tp,p}_\parallel =
\left[ \tilde{A}^{tp}_0+ \tilde{A}^{tp}_1 \left(1-\frac{1}{2}\mu^{tp,p}d\right)
\right] e^{-\mu^{tp,p}d} + \frac{a^{tp,p}_\parallel}{d^4} ,\nonumber\\
&&\mathcal{D}^{tp,t}_\perp =
(A^{tp}_0+ A^{tp}_1) e^{-\mu^{tp,t}d} + \frac{a^{tp,t}_\perp}{d^4} ,\nonumber\\
&&\mathcal{D}^{tp,p}_\perp =
(\tilde{A}^{tp}_0+ \tilde{A}^{tp}_1) e^{-\mu^{tp,p}d} + 
\frac{a^{tp,p}_\perp}{d^4} ,\\
&&\mathcal{D}^{pp,t}_\perp =
(A^{pp}_0+ A^{pp}_1) e^{-\mu^{pp,t}d} + \frac{a^{pp,t}_\perp}{d^4} ,\nonumber\\
&&\mathcal{D}^{pp,p}_\parallel =
\left[ A^{pp}_0+ A^{pp}_1 \left(1-\frac{1}{2}\mu^{pp,p}d\right) \right]
e^{-\mu^{pp,p}d} + \frac{a^{pp,p}_\parallel}{d^4} ,\nonumber
\eeq
where the dependence of the various parameters 
on $B$ is understood and will be discussed in the next
section on the basis of the numerical results obtained by lattice simulations of $N_f=2$ QCD.
Similar investigations
exploring cases of broken Euclidean $SO(4)$ symmetry,
like the finite-temperature case~\cite{DDM2003},
show that the non-perturbative coefficients $A_0$ and $A_1$
are the quantities showing the most significant variation;
however, in principle, both the perturbative and the non-perturbative 
coefficients, as well as the correlation length, might
depend on the particular correlator class and on the value 
of the magnetic field.
The only assumption that
can and will be made apriori consists in the following constraint:
$\tilde{A}^{tp}_0 + \tilde{A}^{tp}_1 = A^{tp}_0 + A^{tp}_1$,
meaning that, at $d=0$, the non-perturbative part  of the correlation
functions belonging to the same ``over-class'', as defined by
(\ref{overclasses}), have the same value.
When $B=0$, the non-perturbative coefficients $A_0$, $A_1$ and $\mu$ no more
depend on he particular correlator class and, moreover,
$a^{tt,t}_\parallel = a^{tp,t}_\parallel = a^{tp,p}_\parallel = a^{pp,p}_\parallel \equiv
a_\parallel$ and
$a^{tt,p}_\perp = a^{tp,t}_\perp = a^{tp,p}_\perp = a^{pp,t}_\perp \equiv a_\perp$,
so that the eight functions in Eq. (\ref{newtradpar}) reduce to the two functions
${\cal D}_\parallel$ and ${\cal D}_\perp$ in Eq. (\ref{param-Dpara-Dperp}), with
$a_\parallel \equiv a_0 - a_1$ and $a_\perp \equiv a_0 + a_1$.

\section{Numerical investigation and discussion}
\label{lattice}

We have considered $N_f = 2$ QCD discretized via 
unimproved rooted staggered fermions and the standard plaquette action
for the pure-gauge sector.
The background magnetic field couples to the quark electric
charges and its introduction corresponds to a modification of the 
Dirac operator: in the continuum an appropriate electromagnetic
gauge field $A_\mu$ must be added to the covariant 
derivative, corresponding to additional
$U(1)$ phases entering the elementary parallel transports
in the discretized lattice version.
For a magnetic field 
$\mathbf{B} = B \mathbf{\hat z}$ the functional integral reads
\beq
Z \equiv \int \mathcal{D}U e^{-S_{G}} 
\det M^{1\over 4} [B,q_u]
\det M^{1\over 4} [B,q_d]
\:,
\label{partfun1}
\eeq
\begin{eqnarray}
M_{i,j} [B,q] &=& a m \delta_{i,j} 
+ {1 \over 2} \sum_{\nu=1}^{4}\eta_{i,\nu} \left(
\vphantom{ U^{\dag}_{i-\hat\nu,\nu} }
u(B,q)_{i,\nu} U_{i,\nu} \delta_{i,j-\hat\nu}
\right. \nonumber \\ && - \left.
u^*(B,q)_{i - \hat\nu,\nu} U^{\dag}_{i-\hat\nu,\nu}\delta_{i,j+\hat\nu} 
\right) \:.
\label{fmatrix1}
\end{eqnarray}
where $S_G$ is the gauge plaquette action,
$q_u = 2|e|/3$ and
$q_d = -|e|/3$ ($|e|$ being the elementary charge) are the quark
electric charges, $i$ and $j$ refer to lattice
sites and $\eta_{i,\nu}$ are the staggered
phases. 

The Abelian gauge field $A_y=B x$ and $A_\mu=0$ for $\mu=t,\,x,\,z$, 
which is a possible choice leading to 
$\mathbf{B} = B \mathbf{\hat z}$,
is discretized on the lattice torus (we assume periodic 
boundary conditions in the spatial directions) as 
\begin{eqnarray}
\label{bfieldy}
&& u^f_{i;\,y}=e^{i a^2 q_f B i_x} \ , \ \ \ \ \ \ 
{u^f_{i;\,x}|}_{i_x=L_x}=e^{-ia^2 q_f L_x B i_y} \ \ \ \ 
\end{eqnarray}
with $u^f_{i;\,\mu}=1$ elsewhere, while
periodicity constraints impose to quantize $B$ as 
follows~\cite{thooft, bound3, wiese}
\beq
|e| B = {6 \pi b}/{(a^2 L_x L_y)} \, ; \ \ \ \ \  b \in \mathbb{Z} \, .
\label{bquant}
\eeq

\begin{figure}[t!]
\includegraphics[width=0.92\columnwidth, clip]{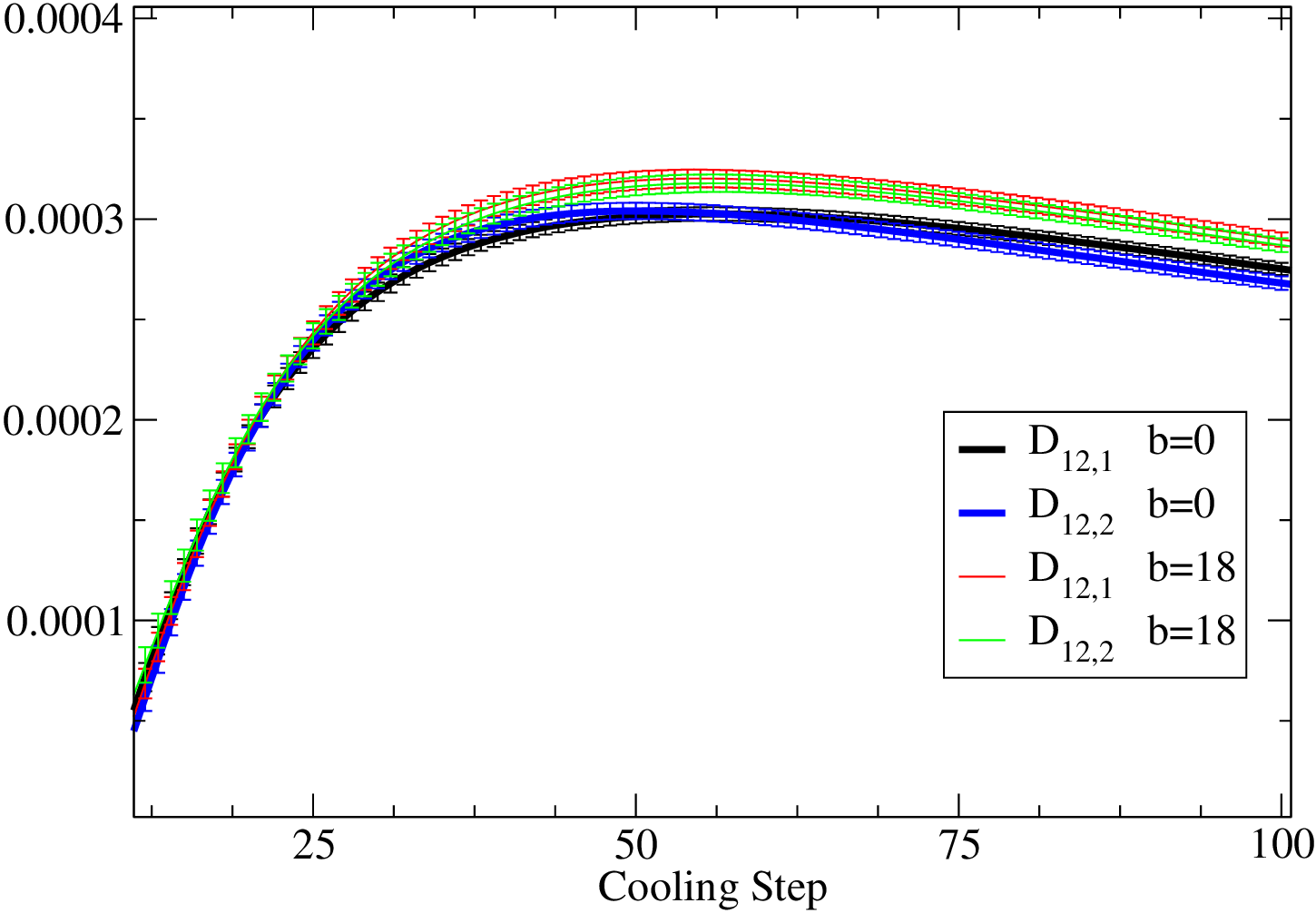}
\caption{Effect of cooling and of the magnetic field on the $\mathcal{D}_{12,1}$ 
and $\mathcal{D}_{12,2}$ correlation functions (evaluated for $d/a=8$), defined in 
Eq.~(\ref{latcorr_measured}). Both correlation functions belong to the 
$\mathcal{D}_\parallel^{tt,t}$
class, defined in Table \ref{table:corrclasses}.}
\label{immu1}
\end{figure}

The correlator in Eq.~(\ref{latcorr_measured}) has been discretized 
through the following lattice observable~\cite{DP1992}
\beq
\mathcal{D}^L_{\mu \nu,\xi}(d) = \left\langle 
Tr\left\{\Omega^\dagger_{\mu\nu}(x) S(x,x+d\hat{\xi})\right. \right.\\ 
\left.\left.   \Omega_{\mu\nu}(x+d\hat{\xi}) S^\dagger(x,x+d\hat{\xi})\right\} 
\right\rangle \nonumber \label{lattice_correlator_form}
\eeq
where $\Omega_{\mu\nu}(x)$ stands for the traceless anti-Hermitian part of the 
corresponding plaquette. In order to remove ultraviolet fluctuations, 
following previous studies of the gauge-field correlators,
a cooling technique has been used~\cite{cooling,Campostrini:1989ts} which,
acting as a diffusion process, smooths out short-distance fluctuations
without touching physics at larger distances: for a correlator 
at a given distance $d$, this shows up as an approximate 
plateau in the dependence
of the correlator on the number of cooling steps, whose
location defines the value of the correlator.
In Fig.~\ref{immu1}, the effect of cooling on 
the correlation function is shown for one particular explored case:  
the value of the correlation 
function is taken at the maximum, and the error is assumed as the independent sum of a 
statistical error and systematic error due to the uncertainty in the determination 
of the plateau, estimated as the difference between the 
value at the maximum and the 
mean of the two neighbouring points at the plateau.

We stress that the prescription adopted in the present study is one among 
other possible definitions of the correlators, which we have chosen
consistently with previous lattice studies of the same quantities.
It will be surely interesting, in the future, to investigate in more details
the issue of the dependence of the correlators on the smoothing procedure.
Let us sketch the main open issues. First, one could adopt a different
smoothing technique, like the so-called {\it gradient flow}~\cite{gradient}:
while cooling and the gradient flow have been shown to provide
equivalent results for the determination of topological 
quantities~\cite{coolingflow,coolflow2}, the situation could be 
different in the present case. Second, the adopted procedure 
implies that a different number of cooling steps is taken
for different correlators; in particular, as noted in previous 
literature on the subject (see, e.g., Ref.~\cite{plb97}), 
the number of cooling steps at which the maximum is reached increases 
approximately quadratically with the correlator distance, as expected
for a diffusion process. Therefore the adopted definition, 
consistently with previous studies, is one in which the regulator
is scaled proportionally to the explored physical distance.
Of course, one could adopt a different definition, in which 
the regulator (i.e. the number of cooling steps) is kept fixed: correlators at larger distances,
for which the maximum is broader and resembles an extended plateau, and 
which are also the ones most sensitive to non-perturbative effects,
do not change dramatically. 
In the present study, we regard such ambiguities as a possibile systematic effect,
related to the very definition of the correlator, which is not included in the reported errors.
However, we note that large part of this systematics is expected to cancel when one considers
the dependence of the correlators on the magnetic field, which is the main
issue considered in the present study. For that reason, 
a detailed comparison of different smoothing procedures is
left to a forthcoming investigation.

Numerical simulations have been performed on a $24^4$ lattice by means of
the Rational Hybrid Monte-Carlo (RHMC)~\cite{Gottlieb:1987mq,Kennedy:1998cu}
algorithm implemented on GPU cards~\cite{gpu}, with statistics 
of $O(10^3)$ molecular dynamics (MD) time 
units
for each $b$ (with $b$ ranging from $0$ to $27$).
The bare parameters have been set to $\beta=5.55$ and $am =0.0125$, 
corresponding to a lattice spacing $a \simeq 0.125$ fm and to a
pseudo-Goldstone pion mass $m_\pi \simeq 480$ MeV \cite{Gottlieb:1992ii}.
The correlators have been measured on about 100 configurations for each 
explored value of $|e|B$, chosen once
every 20 molecular dynamics trajectories. The effects of autocorrelation in the data were 
assessed with a blocking procedure, and appear to be negligible.

\begin{figure}[t!]
\includegraphics[width=0.92\columnwidth, clip]{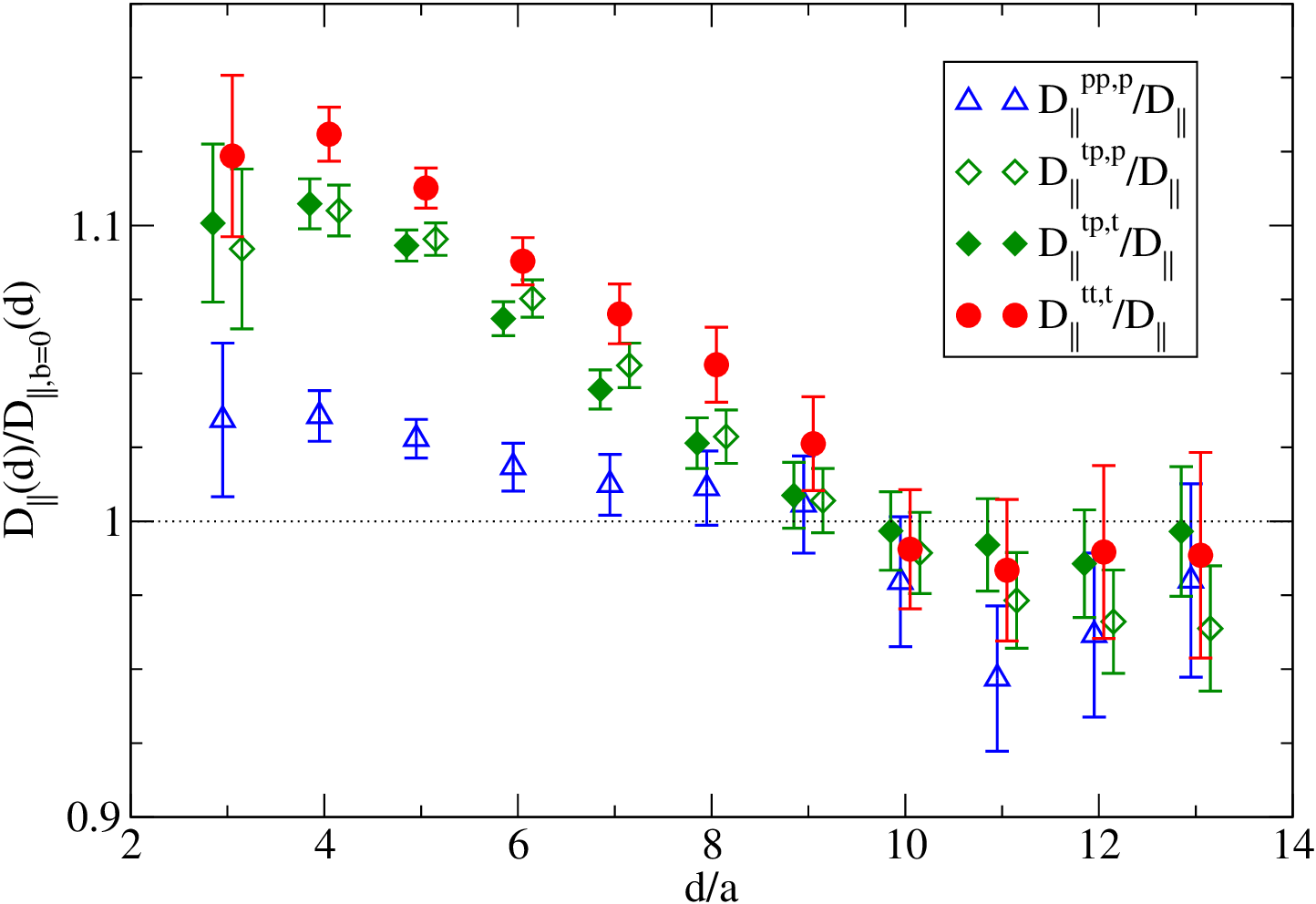}
\caption{Effect of the magnetic field on the \emph{parallel} correlators: the ratio 
$\mathcal{D}^{class}_\parallel(d)/\mathcal{D}_\parallel(d)$ is plotted, where 
$\mathcal{D}^{class}_\parallel(d)$ is measured at $|e|B=1.46$ GeV$^2$ and 
$\mathcal{D}_\parallel(d)$ at $B=0$ . The data points are shifted horizontally for the 
sake of readability.}
\label{Dparratios}
\end{figure}

\begin{figure}[t!]
\includegraphics[width=0.92\columnwidth, clip]{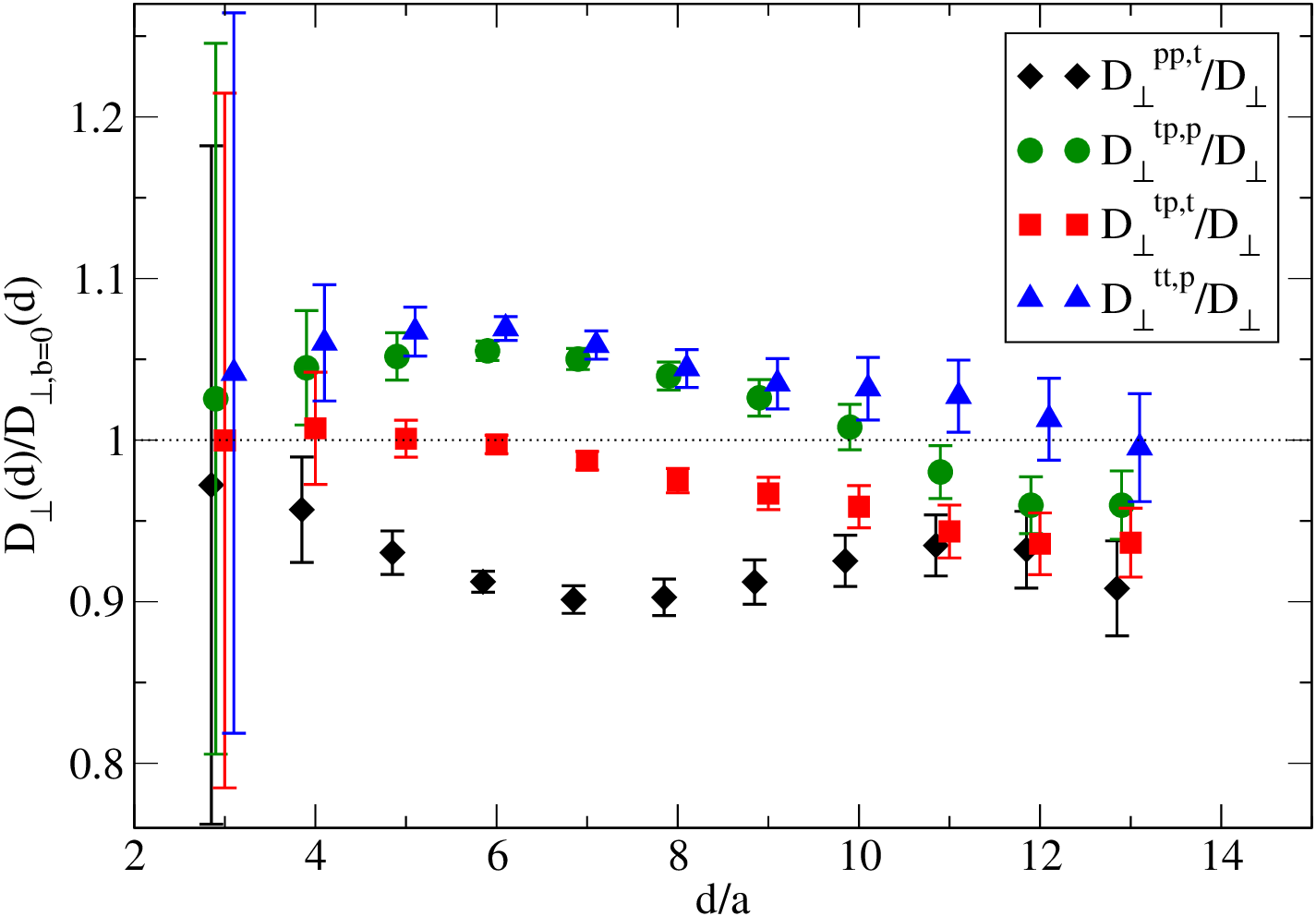}
\caption{Effect of the magnetic field on the \emph{perpendicular} correlators: the 
ratio 
$\mathcal{D}^{class}_\perp(d)/\mathcal{D}_\perp(d)$ is plotted, where 
$\mathcal{D}^{class}_\perp(d)$ is measured at $|e|B=1.46$ GeV$^2$ and 
$\mathcal{D}_\perp(d)$ at $|e|B=0$. The data points are 
shifted horizontally for the sake of readability.}
\label{Dperpratios}
\end{figure}

\subsection{Results and analysis}

To give an example of the dependence of the correlators on the magnetic 
field, 
in Figs. \ref{Dparratios} and \ref{Dperpratios} we plot
results obtained respectively for the 
parallel and for the perpendicular classes, 
as a function of the distance, for the particular case 
$|e|B = 1.46$ GeV$^2$. Correlators are normalized to the values
they take at zero external field. The huge error bars on the points at $d=3a$ 
in Fig.~\ref{Dperpratios} are due to the fact that, for the perpendicular correlators 
at that distance, the flat region surrounding the maximum is considerably narrower 
than for the other distances, and the procedure explained in the previous section to 
assess the systematic uncertainties takes this into account, yielding a larger error 
as one would reasonably expect. This happens both at $|e|B=0$ and at $|e|B=1.46$ GeV$^2$, 
and the resulting effect in the ratios plotted in Fig.~\ref{Dperpratios} is  even larger.

For each value of $|e|B$, we have fitted the correlators with the
parametrization (\ref{newtradpar}), including distances in the range
$3\leqslant d/a \leqslant 8$, thus obtaining an estimate for all 
parameters.
From this first step, it has emerged that the $8$ parameters pertaining to the perturbative 
part of the correlation functions satisfy, within errors, the following equalities: 
\beq
 a^{tt,t}_\parallel \simeq a^{tp,t}_\parallel \simeq a^{tp,p}_\parallel \simeq    a^{pp,p}_\parallel \equiv  a_\parallel & \nonumber \\
 a^{tt,p}_\perp \simeq  a^{tp,t}_\perp  \simeq a^{tp,p}_\perp \simeq  a^{pp,t}_\perp \equiv  a_\perp & \ ,   
\label{pert_assumption}
\eeq
\begin{table}
 \begin{tabular}{|c|c|c|c|c|}
  \hline
  $a_\parallel(B=0)$ & $a^{tt,t}_\parallel$ & $a^{tp,t}_\parallel$ & $a^{tp,p}_\parallel$ &   $a^{pp,p}_\parallel$ \\
  \hline
 0.266(16) & 0.279(14) & 0.277(9)  & 0.272(9) & 0.275(14) \\  
  \hline
  $a_\perp(B=0)$ & $a^{tt,p}_\perp$ & $a^{tp,t}_\perp$     & $a^{tp,p}_\perp$     &   $a^{pp,t}_\perp$ \\  
  \hline
 0.929(16) & 0.94(3) & 0.873(20) & 0.913(23) & 0.88(3) \\  
  \hline
 \end{tabular}
\caption{Values for the perturbative coefficients in Eq.~(\ref{newtradpar}), for $|e|B = 1.46$GeV$^2$. The values of the corresponding parameters at $B=0$ are also reported for comparison.} 
\label{pert_evidence}
\end{table}

as it is possible to see from Table~(\ref{pert_evidence}); moreover {their dependence on $|e|B$ is negligible}. This means that as far 
as the perturbative part of (\ref{newtradpar}) is 
concerned, the parameters introduced in the $|e|B=0$ case are enough to describe our 
data. \\
Driven by this evidence, we have performed a best fit on all measured
correlation functions for $b \leq 18 $, assuming the perturbative parameters
$a_\perp$ and $a_\parallel$ to be independent of $|e|B$, but without making
any {further} assumption about the $|e|B$-dependence of the other parameters.
From this fit we obtain the satisfactory $\chi^2 / n_{dof} = 335 / 322$:
{therefore the parametrization in Eq.~(\ref{newtradpar})
together with the assumptions in Eq.~(\ref{pert_assumption}) 
are assumed in the following
discussion.}

\begin{figure}[t!]
\includegraphics[width=0.92\columnwidth, clip]{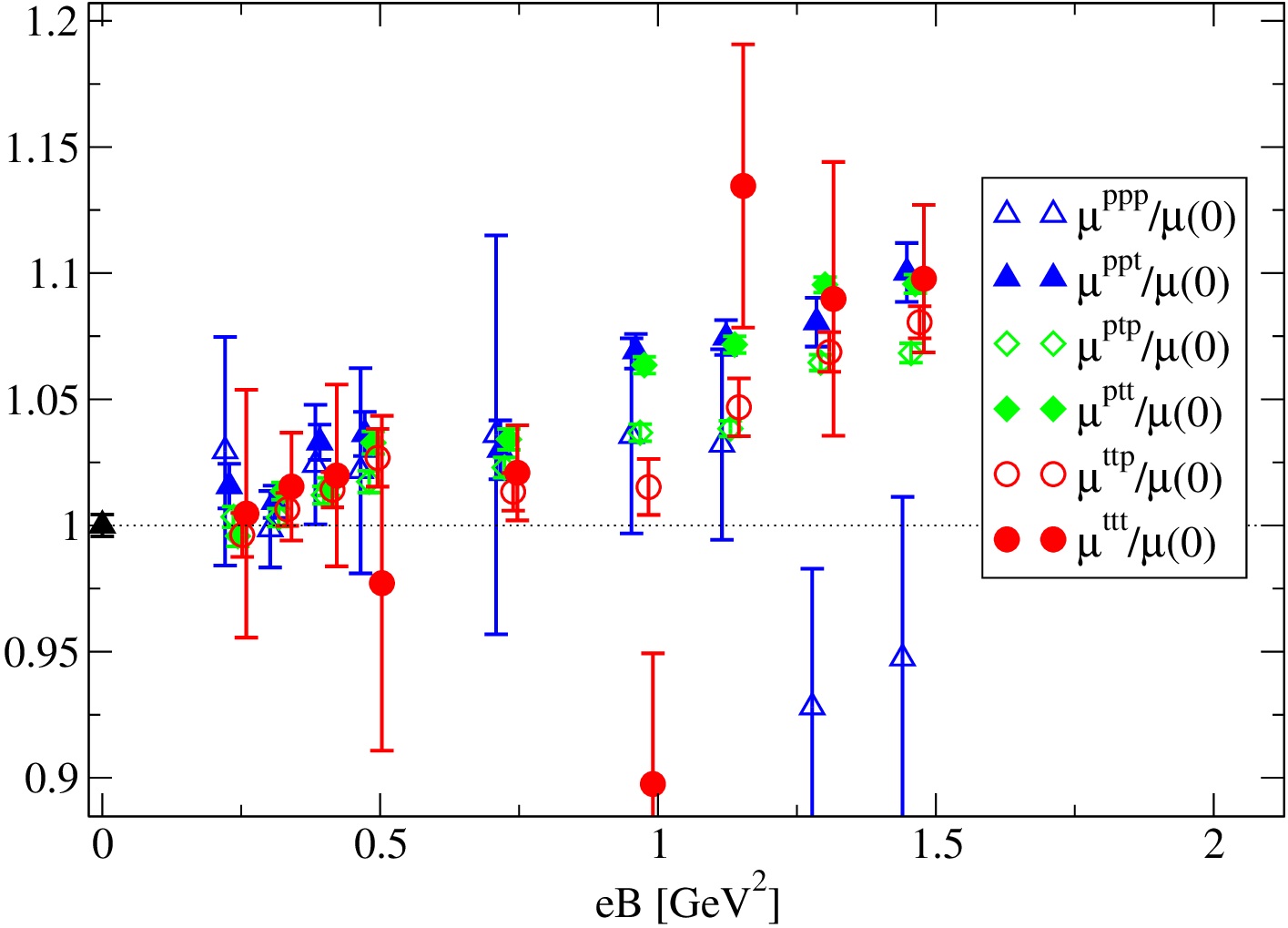}
\caption{Effect of the magnetic field on the (inverse) correlation lengths defined in 
parametrization (\ref{newtradpar}): the ratio $\mu^{class}(|e|B)/\mu(0)$ is plotted.
The value of the (inverse) correlation length for $|e|B=0$ 
is $0.721(3)$ GeV.  The data points are shifted horizontally for the sake 
of readability.}
\label{Lambdas}
\end{figure}

In Fig. \ref{Lambdas}, the inverse of the correlation lengths are plotted as a
function of $|e|B$. It is {not trivial} to give an interpretation of the data: they show a modest 
decrease for most of the correlation lengths, which amounts to about $5-10\%$ for the 
largest values of $|e|B$. We have also performed a fit setting the 6 correlation lengths in 
Eq.~(\ref{newtradpar}) equal, however in this case
 we have obtained $\chi^2 / n_{dof} = 1237 / 358 $, which is not 
satisfactory. We also notice that correlation lengths along the direction perpendicular to 
$B$ (empty symbols in Fig.~\ref{Lambdas}) tend to be slightly smaller than the 
corresponding ones along the parallel direction.

{We have also varied the fit range to assess the part of systematic error 
in our results 
which is related to this choice, exploring the ranges $3 \leqslant d/a \leqslant 7$,  $4 
\leqslant d/a \leqslant 8$ and  $4 \leqslant d/a \leqslant 9$, in addition to  $3 \leqslant 
d/a \leqslant 8$. The estimates of the perturbative parameters $a_\parallel$ and $a_\perp$ 
show a maximum variation around 10\%.
The estimates of the correlation lengths have a maximum variation aroud 5\%:
however, when the ratios to the corresponding values at $|e|B=0$ are considered
(see Fig.~\ref{Lambdas}), the systematic effect is comparable or smaller in
size than the statistical error.
}

{Among the various parameters entering Eq.~(\ref{newtradpar}), the ones
showing the most pronounced variation with $|e| B$ have been the 
non-perturbative coefficients $A_0$ and $A_1$.
That implies a significant dependence on the magnetic field 
of the {\it gluon condensate}, which is discussed in detail
in the next section.}

\subsection{Gluon condensate}
\label{discussion}
\begin{figure}[t!]
\includegraphics[width=0.92\columnwidth, clip]{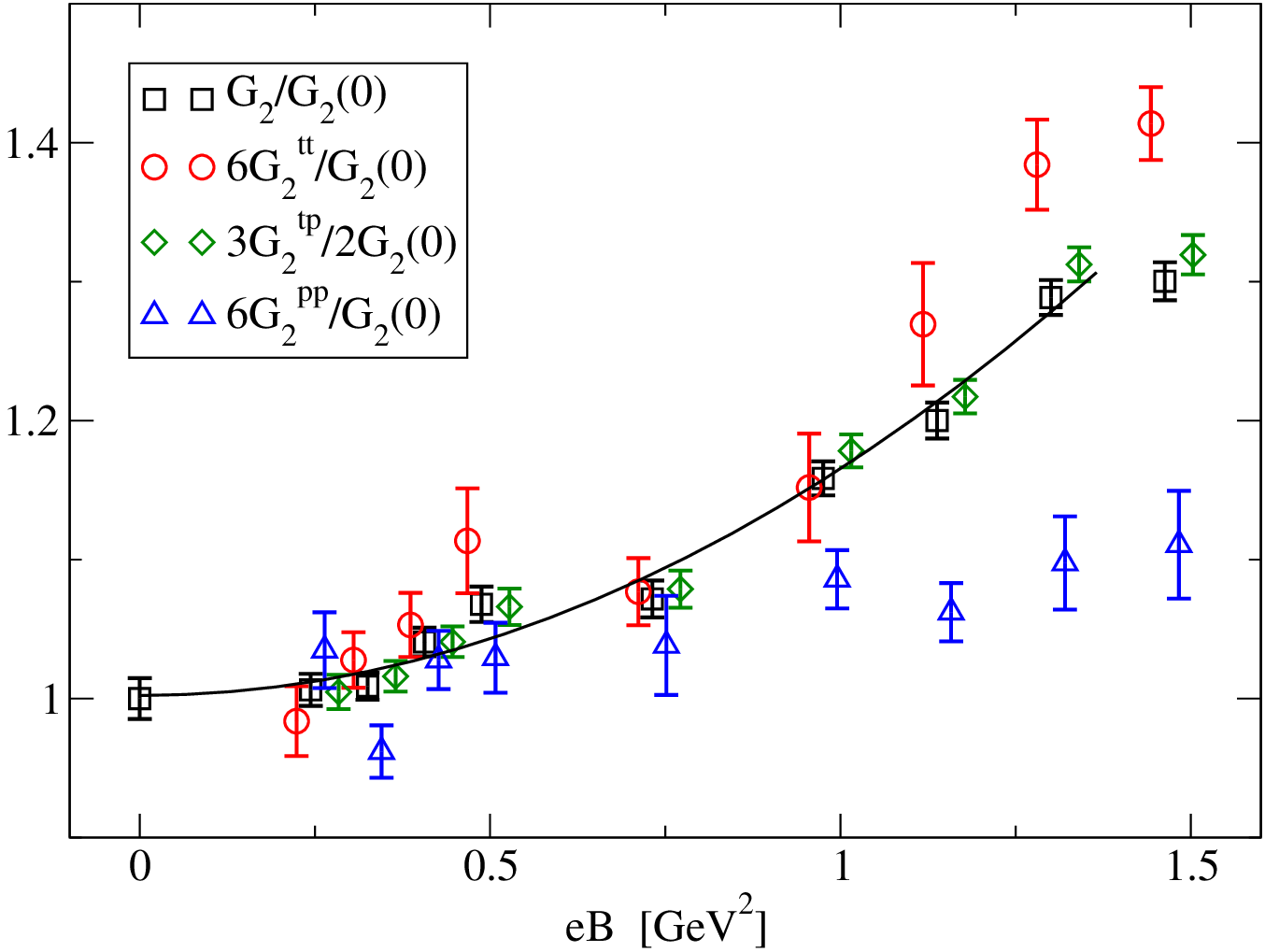}
\caption{Effect of the magnetic field on the gluon condensate $G_2$ obtained 
from the parametrization (\ref{newtradpar}). The effects on the three different contributions 
are plotted along  the total value.  The data points are shifted horizontally for the sake 
of readability.}
\label{g2}
\end{figure}

The {\it gluon condensate} is defined as 
\beq
\label{g2def}
G_2 = \frac{g^2}{4\pi^2} \sum_{\mu \nu, a}
\langle G^a_{\mu\nu} G^a_{\mu\nu} \rangle
\eeq
and is related to the correlator in Eq.~(\ref{defcorr}) 
through an
{\it Operator Product Expansion}. It encodes the main effect of
non-perturbative physics to gluon dynamics and its relevance was first
pointed out in Ref. \cite{SVZ79}.
In the case of $|e|B\neq0$ we can distinguish three contributions, coming from
different sets of plaquettes in the sum in Eq.~(\ref{g2def}), in a fashion
similar to Eq.~(\ref{overclasses}):
\beq
G_2 =  G_2^{tt} + G_2^{tp} + G_2^{pp} \ .
\eeq
For the relation between the gluon-field correlator (\ref{defcorr}) and $G_2$ the reader can 
refer to \cite{plb97} and references therein;
{one obtains $G_2$ from the small distance limit of the 
non-perturbative contributions to the correlator, hence in practice,
following the parametrization in Eq.~(\ref{newtradpar}), we obtain:}
\beq
G_2(B) = \frac{1}{\pi^2} \left[ A^{tt}_0 + A^{tt}_1 + 4 \left(A^{tp}_0 + 
A^{tp}_1 \right) + A^{pp}_0 + A^{pp}_1 \right].
\label{gluonconda}
\eeq
{In Fig. \ref{g2} we report 
the values obtained for $G_2$ as a function of $|e|B$, normalized 
to the value of the condensate obtained for $|e|B = 0$, 
where we obtain {$G_2=3.56(5)\cdot 10^{-2}$ GeV$^4$},
the reported error being just the statistical one.
We have also estimated the systematic error due to the 
choice of the fit 
range, as discussed in the previous subsection: 
the effect on the absolute value of $G_2$ 
at $|e|B=0$ is significant and amounts to 20\% of the total value.
We are therefore in rough agreement, taking also 
into account the unphysical mass spectrum of our discretization,
with the phenomenological estimates of the gluon condensate~\cite{narison}, 
reporting $G_2 \simeq 2.4(1.1) \cdot 10^{-2}$ GeV$^4$.
Instead, when we consider the ratios of condensate to 
the $|e|B = 0$ value, which are reported in 
Fig.~\ref{g2},
the systematic error turns out to be negligible
as compared to the statistical one.}

{We notice that $G_2$ grows as a function of $|e|B$, the increase being
of the order of $25\%$ for the largest value of $|e|B$ explored.
In 
the same figure we also report the relative increases in the $G_2^{tt}$,$G_2^{tp}$ and 
$G_2^{pp}$ terms. We see that the $tt$ term is the most affected by the magnetic field, 
whereas the $pp$ contribution shows a really modest dependence on $|e|B$.
In Fig. \ref{g2}, the best fit  with a quadratic 
function 
\beq
\frac{G_2(|e|B)}{G_2(0)} = 1 + K (|e|B)^2
\eeq
is also plotted.
We obtain $K=0.164(7)\ \textnormal{GeV}^{-4}$ and
 $\chi^2/n_{dof} = 1.52$, 
excluding the point at $|e|B=1.46$ GeV$^2$.
}
{ An increase  of the chromomagnetic gluon condensate with $|e|B$ 
has also been found in \citep{ozaki}, which is  in qualitative agreement 
with the result presented here.  A similar behaviour for $G_2$ has been also 
predicted making use of QCD sum rules \cite{Ayala:2015qwa}.}

\section{Conclusions}
\label{finally}

In this study we have explored the effects of a magnetic background field
on the gauge-invariant two-point correlators of the gauge-field strength. 
Electromagnetic fields are coupled directly to quark fields: however, as for
other pure-gauge observables, we did expect and we have indeed observed a
non-trivial effect which can be interpreted in terms of non-perturbative
quark-loop contributions.
We have discussed the residual symmetries of the Lorentz group in the 
presence of a constant and uniform magnetic 
background, and how it affects the general structure of the correlators. We 
have then presented the numerical results of an exploratory lattice
study performed for $N_f = 2$ QCD discretized via rooted staggered fermions.

Our results can be summarized as follows. We have evidenced a significant effect of the magnetic field on the correlation functions (see Figs.~\ref{Dparratios} and \ref{Dperpratios}).
The short-distance, perturbative part
of the correlators is practically unaffected by the presence of the magnetic
background, while significant effects are observed for the non-perturbative 
part. In particular, one observes a mild effect on the non-perturbative correlation
lengths, with a general tendency for a decrease of the correlation lengths as a 
function of $|e|B$, which is slightly more visible for correlators
in the directions orthogonal to $|e|B$. 

A larger effect, and a more significant anisotropy, is observed for 
the coefficients of the non-perturbative terms, which can be directly 
related to the gluon condensate (see Eq.~(\ref{gluonconda})). Due to the explicit
Lorentz symmetry breaking caused by the magnetic field, we can distinguish among
3 different contributions to the gluon condensate. An analysis based on Eq.~(\ref{gluonconda}) shows that each term has a different 
behaviour as a function of the magnetic field (see Fig.~\ref{g2}).
Starting from that, we have observed that the gluon condensate itself increases 
as a function of $B$, with the increase being of the order of 20\%
for $|e|B \sim 1$ GeV$^2$. Relative differences between the different
contributions are of the same order of magnitude, meaning that 
anisotropies induced by $B$ are significant and comparable
to those observed in other pure-gauge quantities (see, e.g., Ref.~\cite{struzzo}). 
The increase of the gluon condensate provides evidence of the phenomenon
known as {\it gluon catalysis}, which had been previously observed
based on the magnetic-field effects on plaquette expectation values 
\cite{Bali:2013esa, Ilgenfritz:2012fw,Ilgenfritz:2013ara}.
We notice that the overall effect on the correlators, in particular
regarding the changes in the perturbative and non-perturbative parts, 
is similar to what has been observed in other setups where Lorentz symmetry 
is explicitly broken by external parameters, like for QCD at finite temperature
\cite{DDM2003}.  

In the future, we plan to repeat the present exploratory study by adopting
a discretization of QCD at the physical point, i.e. with quark masses
tuned at their phenomenological values, 
{and by extending the investigation to other
gauge invariant correlation functions, like those involving
quark fields~\cite{quarkcorr}}. It will be also interesting to study,
within the Stochastic Vacuum Model, which takes the correlators as 
an input, the effect of the magnetic background field on the static 
quark-antiquark potential, and in particular on the string tension, in order
to obtain an indirect confirmation of the anisotropy of the potential
which has been already observed by direct lattice measurements~\cite{struzzo}.
Finally, as we have already observed in Section~\ref{lattice}, in this work
we have limited ourselves to the smoothing technique ({\it cooling})
adopted in previous lattice studies for the measure of the correlators,
but one could also adopt the so-called {\it gradient flow}~\cite{gradient}
as a regulator, and consider a different prescription for fixing the amount
of smoothing: we postpone a careful comparison of different smoothing
procedures for these particular observables to a forthcoming investigation.

\acknowledgments

FN acknowledges financial support from the INFN SUMA project.
We thank the Scientific Computing Centre at INFN-Pisa and INFN-Genoa for 
providing computing resources.

\end{document}